\documentclass{jnmp}
\usepackage{amsmath}
\newtheorem*{theorem}{Theorem}

\begin{document}
\renewcommand{\evenhead}{A M Perelomov}
\renewcommand{\oddhead}{On Frequencies of Small Oscillations
of Some Dynamical Systems}

\thispagestyle{empty}
\FirstPageHead{10}{1}{2003}{\pageref{perelomov-firstpage}--
\pageref{perelomov-lastpage}}{Article}
\copyrightnote{2003}{A M Perelomov}

\Name{On Frequencies of Small Oscillations\\
 of Some Dynamical Systems\\ Associated  with Root Systems
 \footnote{math-ph/0205039}}
\label{perelomov-firstpage}

\Author{A M PERELOMOV}
\Address{Departamento de F\'{\i}sica, Facultad de Ciencias,
Universidad de Oviedo, Oviedo, Spain\\
On leave of absence from Institute for Theoretical and Experimental
Physics,\\ 117259 Moscow, Russia\\
E-mail: perelomo@dftuz.unizar.es}
\Date{Received May 29, 2002; Accepted July 07, 2002}

\begin{abstract}\noindent
In the paper by F Calogero and the author [{\it Commun. Math. Phys.}
{\bf 59} (1978), 109--116] the formula
for frequencies of small oscillations of the Sutherland system
($A_l$ case) was found. In the present note the generalization of this
formula for the case of arbitrary root system is given.
\end{abstract}

\section{Introduction}
In this note, we consider the classical Hamiltonian
systems associated to the root systems, with the potential
$v(q)=g^2\,\sin ^{-2}q$. Such systems were introduced in \cite{op1} as
the generalization of the Calogero--Sutherland  systems \cite{ca1,su}.

It is known that these systems have the equilibrium configurations
which are distinguished by many remarkable features (see papers
\cite{abcop,bms,ca2,ca3, cp1,cp2,cs,kps,mop,op2,pe1} and  books
\cite{ca4,pe2}). In the simplest case ($A_l$ case with the
potential $v(q)=g^2\,\sin ^{-2}q$), the explicit expression for
frequencies of small oscillations near equilibrium was found in
the paper~\cite{cp1}. For the case of the potential
$v(q)=q^2+g^2q^{-2}$, these frequencies are known for the case of
arbitrary root system and they are proportional to the degrees of
basic invariants of the Weyl group of the corresponding root
system \cite{pe1}.

The main result of this note is the explicit formula for the
frequencies of such oscillations for the trigonometric case.
Namely, these frequencies
are proportional to the coefficients $r_j(g)$ in the expansion of the sum
of positive roots in the simple roots $\{ \alpha _j\}$:
\begin{equation}
2\rho (g) =\sum _{\alpha \in R^+} g_{\alpha }\alpha =
\sum _{j=1}^lr_j(g) \alpha _j.
\end{equation}
This is the generalization of the old result by Calogero and
author~\cite{cp1} (for numerical calculations of such type quantities
see~\cite{cs}).

As a byproduct of the calculations at $g_\alpha =1$, we discovered the
identity for the root
systems which we cannot find in the literature, namely,
\begin{equation}
\prod _{j=1}^l(d _j-1)d_j=z\cdot\prod _{k=1}^l r_k .
\end{equation}
Here $\{d_j\}$ are the degrees of the basic invariants of the root
systems, and $z=z( G )$ is the dimension of the center of the simple compact
simply-connected Lie group $G$ corresponding to the root system $R$.

\section{General description}
The systems under consideration were introduced in \cite{op1} and they are
described by the Hamiltonian (for more details, see \cite{op2} and \cite{pe2})
\begin{equation}
H=\frac{1}{2} p^2 + U(q),\qquad p^2=(p,p)=\sum _{j=1}^l p_j^2,
 \end{equation}
where $p=(p_1,\ldots,p_l)$, and $q=(q_1,\ldots,q_l)$ are the
momentum vector and the coordinate one in the $l$-dimensional
vector space $V\sim {\mathbb R}^l$ with the standard scalar
product $(\cdot ,\cdot)$. The potential $U(q)$ is constructed by
means of the certain system of vectors $R=\{\alpha\}=R^+\cup R^-$
in $V$, so-called root system.
\begin{equation}
U=\sum \limits_{\alpha \in R^+}g_\alpha ^2 v(q_\alpha ),\qquad
q_\alpha =(\alpha ,q).
\end{equation}
The constants $g _\alpha$ satisfy the condition $g_\alpha =g_\beta $ if
$(\alpha ,\alpha )=(\beta, \beta ).$ Such systems are
completely integrable for $v(q)$ of five types. Here we consider
only the case $v(q)=\sin ^{-2}q$.

Note that these systems are the generalization of the Calogero--Sutherland
systems \cite{ca1} and \cite{su}, for which $R^+=\{e_i-e_j,\,\,i<j\}$
and $\{e_j\}$, $j=1,\ldots,l+1$ is the standard basis in ${\mathbb R}^{l+1}$.

\section{Root systems}
We give here the basic definitions. For more details, see
\cite{hu,ma2} and \cite{ov}. Let $V$ be the $l$-dimensional real
vector space with a standard scalar product $(\cdot ,\cdot )$,
$(\alpha ,\beta )=\sum \alpha _j\beta _j$, and let $s_{\alpha}$ be
the reflection in the hyperplane through the origin orthogonal to
the vector $\alpha$
\begin{equation}
s_{\alpha} q = q - (q,\alpha ^\lor )\,\alpha ,\qquad \alpha ^\lor =
\frac{2}{(\alpha ,\alpha )}\,\alpha .\end{equation}
Consider a finite set of nonzero vectors $R=\{\alpha\}$
generating $V$ and satisfying the following conditions:

\begin{enumerate}\itemsep0mm
\item[1.] For any $\alpha\in R$, the reflection $s_{\alpha}$ conserves
$R :s_{\alpha}R = R$.
\item[2.] For all $\alpha,\beta \in R$, we have $(\alpha ^\lor ,\beta )\in
{\mathbb Z}.$
\end{enumerate}

The set $\{s_\alpha \}$ generates the finite group $W(R)$ -- the
Weyl group of $R$. The root system~$R$ is called  {\em reduced
system} if only vectors in $R$ collinear to $\alpha $ are $\pm
\alpha $. Let us choose the hyperplane which does not contain the
root. Then we have $R=R^+\bigcup R^-$, and $R^+$ is the set of
positive roots. In $R^+$ there is the basis $\{ \alpha _1,\ldots
,\alpha _l\}$ (the set of simple roots) such that $\alpha =\sum _j
\,n_j\alpha _j$, $n_j\geq 0$ for any $\alpha \in R^+$. The root
system $R$ is called {\em irreducible system} if it can not be the
union of two non-empty subsets $R_1$ and $R_2$ which are
orthogonal each other.

Let $\{\alpha _1,\ldots ,\alpha _l\}$ be the set of simple roots in $R$,
$R^{+}$ be the set of positive roots, and $\{\lambda _j\}$ be a dual basis
or the weight basis: $(\lambda _j, \alpha _k)=\delta _{jk}$.

Let $Q$ be the root lattice, and $Q^+$ be the cone of positive roots
\begin{equation}
Q= \left \{ \beta :  \beta =\sum\limits _{j=1}^l m_j\alpha _j,
m_j\in {\mathbb Z}\right \};\qquad
Q^+=\left \{ \gamma :  \gamma = \sum \limits_{j=1}^l n_j\alpha _j,
\ n_j\in {\mathbb N} \right \}.
\end{equation}
Let $P$ be the weight lattice, and $P^+$ be the cone of dominant weights
\begin{equation}
P=\left \{ \lambda :  \lambda = \sum\limits_{j=1}^l m_j\lambda _j,
\ m_j\in {\mathbb Z} \right \};\qquad
P^+ = \left \{\mu : \mu =\sum\limits_{j=1}^l n_j
\lambda _j,\ n_j\in {\mathbb N} \right \}.
\end{equation}

Following \cite{ma2}, we define a partial order on $P$ as
follows: $\lambda \geq \mu $ if and only if $\lambda -\mu \in Q^+$
(or $(\lambda ,\lambda _j)\geq (\mu ,\lambda _j)$ for all $j=1,\ldots ,
l$). The set of linear combinations over ${\mathbb R}$ of the functions
$f_\lambda (q)= \exp \{2i(\lambda ,q)\}$, $\lambda \in P$, $q\in V$
may be considered as the group algebra~$A$ over $\mathbb R$ of the free
abelian group $P$. For any $\lambda \in P$, let us denote
$e^\lambda \sim f_\lambda (q)$ as the corresponding element of $A$,
so that $e^\lambda e^\mu =e^{\lambda +\mu }$, $(e^\lambda )^{-1}=
e^{-\lambda }$, and $e^0=1$, the identity element of $A$. Then
$e^\lambda $, $\lambda \in P$ form the $\mathbb R$-basis of $A$.

The Weyl group $W(R)$ acts on $P$ and hence also on $A:s(e^\lambda )=
e^{s\lambda }$ for $s\in W$ and $\lambda \in P$.
We denote as $A ^W$ the subalgebra of $W$-invariant elements of $A$.

It is known that $A^W$ is the algebra free generated by basic invariants of
degrees $d_1,\ldots,d_l$. We introduce also a $g$-deformed vector
\[
\rho (g)=\frac12 \sum _{\alpha \in R^+} g_\alpha \alpha .
\]
Let $\delta $ be the highest root, i.e.\ the positive root such that
$\delta \geq\alpha$ for all $\alpha \in R^+$.

\section{Results}\resetfootnoterule
Let us consider two classical systems characterized by the Hamiltonians
\begin{equation}
H_m=\frac12\,p^2+U_m(q),\qquad m=1,2 \end{equation}
with
\begin{gather}
U_1(q) = -\sum _{\alpha \in R^+} g_\alpha \log |\sin (q_\alpha )|,\\
U_2(q) = \sum _{\alpha \in R^+} g_\alpha ^2 \sin ^{-2}(q_\alpha ),
\qquad  q_\alpha =(\alpha ,q),\quad g_{\alpha}=g_{\beta}\quad {\rm if}\quad
 (\alpha ,\alpha)=(\beta ,\beta).
\end{gather}

The configuration space of such system is the Weyl alcove
\begin{equation}
\{ q\,|\,(q,\alpha _j)>0,\  j=1,\ldots ,l,\  (q,\delta )<\pi \},
\end{equation}
where $\{\alpha _j\}$ is the set of the simple roots, $\delta $ is
the highest root, and as it well known the potential energies
$U_1$ and $U_2$ have an  isolated minimum at the point $\bar q$
inside the alcove\footnote{Let us note that the first system is
the generalization of the so-called ``Dyson system''
\cite{dy1,dy2}.}. Let us denote this minimum as $U_1^{(0)}$ and
$U_2^{(0)}$, $U_m^{(0)}= U_m(\bar q)$. Then near this point we
have
\begin{equation}
U_m(q)\sim U_m^{(0)}+\frac12\,a_{jk}^{(m)} \xi _j\xi _k,\qquad \xi _j=q_j-
{\bar q}_j.
\end{equation}
The matrices $a^{(m)}$ are positive definite and as it is known~\cite{pe1},
\begin{equation}
a^{(2)}=c\big( a^{(1)}\big) ^2,\qquad c=\mbox{const} .
\end{equation}
So, we need to know just one of them. The frequencies of small
oscillations $\omega _j^{(m)}$ near the equilibrium configuration
are square roots of eigenvalues of these matrices and we have
$\omega^{(2)}_j=c\big( \omega^{(1)}_j\big) ^2$. The eigenvalues of
these matrices were known before only for $A_l$ case
(see~\cite{cp1}). Here we give the explicit expression for them
for the general case.

\begin{theorem}
Eigenvalues of the matrix $a^{(1)}$ are
equal to the quantities $\{2\, r_j (g)\}$, $j=1,\ldots\! ,l$,
where $\{ r_j (g)\}$ are the coefficients in the expansion
\begin{equation}
2\rho (g) = \sum _{\alpha\in R^+} g_{\alpha}\alpha =
\sum _{j=1}^{l} r_j(g)\alpha _j.
\end{equation}
\end{theorem}

\begin{proof}
Note that the function $U_1(q)$ is equal to $-\,\log \left|
\Psi _0^\kappa \right| $, where $\Psi _0^\kappa $ is the solution of
the Schr\"odinger equation~\cite{op3}
\begin{gather}
\left( \frac12 {\hat p}^2+U_2(q)\right) \Psi _0^\kappa =E_0(\kappa )
\Psi _0^\kappa ,\qquad  g_{\alpha}^2=\kappa _{\alpha}(\kappa _{\alpha}-1),
\quad {\hat p}_j=-i\,\frac{\partial }{\partial q_j}, \\
\Psi _0^\kappa = \prod _{\alpha \in R^+} (\sin q_{\alpha})^{\kappa _\alpha}.
\end{gather}
The function $\Psi _0^\kappa (q)$ is positive inside the Weyl
alcove and has a maximum at point $q={\bar q}$. Let $\kappa
_\alpha =s \tilde \kappa_\alpha $ and let  us consider the limit
as $s \to \infty $.

Then the function
\begin{equation}
\Phi _0^{\tilde \kappa} (\xi )=\lim _{s \to \infty }  \frac{\Psi _0^\kappa
\left(\bar q+\xi/\sqrt{s}\right) }{\Psi _0^\kappa (\bar q)},
\qquad \kappa _\alpha =s \tilde \kappa_\alpha
\end{equation}
takes the Gaussian form\footnote{In other words at the limit
$\kappa_{\alpha}\to\infty$ $(s\to\infty )$ our quantum system
becomes the oscillatory one for which the correspondence between
classical and quantum systems is well known. Note that at this
limit wave functions go to oscillatory ones. For example, for the
case of one degree of freedom we have
\[
\lim _{s\to\infty} s^{(-n/2)} P_n^{(\alpha,\alpha)}(x/\sqrt{s})
=(2^n n!)^{-1} H_n(x) ,
\]
 where $P_n^{(\alpha,\beta)}(x)$ and $H_n(x)$
are Jacobi and Hermite polynomials, correspondingly. For the
multidimensional case, analogous formulae give the multivariable
Hermite polynomials.}
\begin{equation}
\Phi _0^{\tilde \kappa }(\xi )=\exp
\left(-\frac12\,a_{jk}\xi _j\xi _k\right) .
\end{equation}

From other side, at this limit the Schr\"odinger equation
\begin{equation}
\left( \frac{1}{2}\,{\hat p}^2 +U_2(q)\right) \Psi _{\boldsymbol{m}}
=E_{\boldsymbol{m}}
\Psi _{\boldsymbol{m}},\qquad \boldsymbol{m}=(m_1,\ldots ,m_l)
\end{equation}
takes the form
\begin{equation}
\left( -\Delta _\xi +\frac12\sum _{j,k=1}^l b_{j,k}\xi _j\xi _k\right)
\Phi _{\boldsymbol{m}} = \tilde E_{\boldsymbol{m}} \Phi _{\boldsymbol{m}},
\end{equation}
where
\begin{equation} b=a^2,
\qquad \tilde E_{\boldsymbol{m}}=\tilde E_0+\sum _{j=1}^l \omega _j m_j,
\end{equation}
and $\omega _j$ are the eigenvalues of the matrix $a$. Note that
the spectrum of the quantum problem
\[
H\Psi _{\boldsymbol{m}}=E_{\boldsymbol{m}} \Psi _{\boldsymbol{m}}
\]
is given by the formula
\begin{equation}
E_{\boldsymbol{m}}=\sum _{i,j=1}^l (\lambda _i,\lambda _j) m_i\,m_j +
2\sum _{j=1}^l(\lambda _j,\rho (\kappa) )m_j +E_0(\kappa).
\end{equation}
At $s \to \infty $ we obtain
\begin{equation}
\tilde E_{\boldsymbol{m}}=\lim _{s\to\infty}
\left( s^{-1} E_{\boldsymbol{m}}(\kappa)\right) =
\tilde E_0+2\sum _{j=1}^l (\lambda _j,\rho (\tilde \kappa))m_j.
\end{equation}

Comparing (21) and (23), we obtain the main formula
\begin{equation}
\big(\omega ^{(1)}_j\big)^2=2 r_j (\tilde{\kappa})=
2(\lambda _j,\rho (\tilde \kappa)).
 \end{equation}
So, the classical result is obtained from the quantum result as
$s\to \infty $. \end{proof}

In conclusion, we give  the explicit expression for the maximum value of the
function $\Psi _0 ^{\kappa} (q)$.

This formula was given by I~Macdonald as an conjecture~\cite{ma1} that
was proved later by E~Opdam~\cite{op}. For the simplest case
($\kappa _{\alpha}=1$ for all $\alpha$),  it has the form
\begin{equation}
\left| \Psi _0^1 (\bar q)\right| ^2=\left\{\prod _{\alpha \in R^+}
(\sin {\bar q}_\alpha )\right\}^2=\frac{|W|}{2^{|R|}} \prod
_{j=1}^l\left( \frac{d_j}{d_{j}-1}\right) ^{d_j-1},
\end{equation}
where $|W|$ is the order of $W$, and $|R|$ is the number of roots.

\subsection*{Acknowledgements}
I am grateful to Prof. R~Sasaki who raised the question on small
oscillations and to the Physics Department of Oviedo University for the
hospitality.

\label{perelomov-lastpage}


\begin{thebibliography}{99}
\small
\bibitem{abcop} Ahmed S, Bruschi M, Calogero F, Olshanetsky
M~A and Perelomov A~M, Properties of the Zeros of the Classical
Polynomials and of Bessel Functions, {\it Nuovo Cim.} {\bf 49} (1979),
173--199.

\bibitem{bms} Bordner A~J, Manton N~S and Sasaki R,
Calogero--Moser Models V:  Supersymmetry, and Quantum Lax Pairs,
{\it Prog. Theor. Phys.} {\bf 103} (2000), 463--487; hep-th/9910033.

\bibitem{ca1}  Calogero F, Solution of the One-Dimensional
$N$-Body Problem with Quadratic and/or Inversely Quadratic Pair Potentials,
{\it J. Math. Phys.} {\bf 12} (1971),  419--436.

\bibitem{ca2} Calogero F, On the Zeros of the Classical
Polynomials, {\it Lett. Nuovo Cim.} {\bf 19} (1977), 505--508.

\bibitem{ca3} Calogero F, Equilibrium Configuration of the
One-Dimensional $n$-Body Problem with Quadratic and Inversely Quadratic Pair
Potentials, {\it Lett. Nuovo Cim.} {\bf 20} (1977),  251--253.

\bibitem{ca4} Calogero F, Classical Many-Body Problems in
One-Two- and Three-Dimensional Space Amenable to Exact Treatments (Solvable
and/or Linearizable), Springer-Verlag, 2001.

\bibitem{cp1} Calogero F and Perelomov A~M, Properties
of Certain  Matrices Related to the Equilibrium Configuration of
One-Dimensional
Many-Body Problems with Pair Potentials $V_{1}=-\log|\sin x|$ and
$V_{2}=1/\sin^2x$, {\it Commun. Math. Phys.} {\bf 59} (1978), 109--116.

\bibitem{cp2} Calogero F and Perelomov A~M, Some Diophantine
Relations Involving Circular Functions of Rational Angles,
{\it Lin. Alg. Appl.} {\bf 25} (1979),  91--94.

\bibitem{cs} Corrigan E and Sasaki R, Quantum VS Classical
Integrability in Calogero--Moser Systems, 2002, hep-th/0204039.

\bibitem{dy1} Dyson F~J, Statistical Theory of the Energy
Levels of Complex Systems, I, II, III, {\it J. Math. Phys.}
{\bf 3} (1962), 140--156, 157--165, 166--175.

\bibitem{dy2} Dyson F~J, A Brownian Motion Model for the Eigenvalues
of a Random Matrix,  {\it J. Math. Phys.}  {\bf 3} (1962),  1191--1198.

\bibitem{hu} Humphreys J~E, Introduction to Lie Algebras and
Representation Theory, Springer, New York, 1972.

\bibitem{kps} Kogan Ya~I, Perelomov A~M and Semenoff G,
Charge Distribution in Two-Dimensional Electrostatics, {\it Phys. Rev.}
{\bf B45},  Nr.~20 (1992), 12084--12087.

\bibitem{ma1} Macdonald I, Some Conjectures for Root
Systems,  {\it SIAM J. Math. Anal.} {\bf 13} (1982),  988--1007.

\bibitem{ma2} Macdonald I, Orthogonal Polynomials Associated
with Root Systems, Preprint, 1987.

\bibitem{ma3}  Macdonald I, Symmetric Functions and Hall
Polynomials, Oxford Univ. Press, 1995.

\bibitem{mop} Mikhailov A~V, Olshanetsky M~A and Perelomov A~M,
Two-Dimensional Generalized Toda Lattice, {\it Commun. Math. Phys.}
{\bf 79} (1981), 473--488.

\bibitem{op1} Olshanetsky M~A and Perelomov A~M, Completely
Integrable Hamiltonian Systems Connected with Semisimple Lie Algebras,
{\it Invent. Math.} {\bf 37} (1976), 93--108.

\bibitem{op2} Olshanetsky M~A and Perelomov A~M, Classical
Integrable Finite-Dimensional Systems Related to Lie Algebras,
{\it Phys. Reps.}  {\bf 71} (1981), 314--400.

\bibitem{op3} Olshanetsky M~A and Perelomov A~M, Quantum
Integrable Systems Related to Lie Algebras, {\it Phys. Reps.} {\bf 94}
(1983), 313--404.

\bibitem{ov} Onishchik A~L and Vinberg E~B, Lie Groups and
Algebraic Groups, Springer-Verlag, Berlin--Heidelberg, 1990.

\bibitem{op} Opdam E, Some Applications of Hypergeometric
Shift Operators, {\it Invent. Math.} {\bf 98} (1989), 1--18.

\bibitem{pe1} Perelomov A~M, Equilibrium Configurations and
Small Oscillations of Some Dynamical Systems, {\it Ann. Inst. H. Poincar\'e}
{\bf A28} (1978),  407--415.

\bibitem{pe2} Perelomov A~M, Integrable Systems of Classical
Mechanics and Lie Algebras. I, Birkhauser, 1990.

\bibitem{su} Sutherland B, Exact Results for a Quantum
Many-Body Problem in One-Dimension. II, {\it Phys. Rev.} {\bf A5} (1972),
1372--1376.
\end{thebibliography}
\end{document}